\documentclass[aps,prb,twocolumn,superscriptaddress,showpacs,amsmath,amssymb,
footinbib,longbibliography,10pt]{revtex4-2}
\usepackage[utf8]{inputenc}
\usepackage{bm}
\usepackage[usenames]{color}
\usepackage{xcolor}
\usepackage{multirow}
\usepackage{amssymb}
\usepackage{amsbsy}
\usepackage{amsmath}
\usepackage{stmaryrd}
\usepackage{graphicx}
\usepackage{epsfig}
\usepackage{placeins}
\usepackage[normalem]{ulem}
\usepackage{bbold}
\usepackage{bbm}
\usepackage{braket}
\usepackage{tikz}
\usetikzlibrary{calc}
\usetikzlibrary{patterns}
\usepackage[colorlinks,linkcolor=blue,citecolor=blue,urlcolor=blue]{hyperref}

\definecolor{mycolor}{RGB}{0,100,0}

\makeatletter

\setcitestyle{numbers,square,comma,sort&compress}

\renewcommand{\eqref}[1]{Eq.\hspace{2pt}\protect(\ref{#1})}
\definecolor{palette1}{HTML}{A8216B}
\definecolor{palette2}{HTML}{F1184C}
\definecolor{palette3}{HTML}{F36943}
\definecolor{palette4}{HTML}{F7DC66}
\definecolor{palette5}{HTML}{2E9599}

\usepackage{scalerel}
\usepackage{tikz}
\usetikzlibrary{calc}
\usetikzlibrary{patterns}
\usetikzlibrary{svg.path}
\definecolor{orcidlogocol}{HTML}{A6CE39}
\tikzset{
  orcidlogo/.pic={
    \fill[orcidlogocol]
svg{M256,128c0,70.7-57.3,128-128,128C57.3,256,0,198.7,0,128C0,57.3,57.3,0,128,
0C198.7,0,256,57.3,256,128z};
    \fill[white] svg{M86.3,186.2H70.9V79.1h15.4v48.4V186.2z}

svg{M108.9,79.1h41.6c39.6,0,57,28.3,57,53.6c0,27.5-21.5,53.6-56.8,53.6h-41.8V79.
1z
M124.3,172.4h24.5c34.9,0,42.9-26.5,42.9-39.7c0-21.5-13.7-39.7-43.7-39.7h-23.
7V172.4z}

svg{M88.7,56.8c0,5.5-4.5,10.1-10.1,10.1c-5.6,0-10.1-4.6-10.1-10.1c0-5.6,4.5-10.1
,10.1-10.1C84.2,46.7,88.7,51.3,88.7,56.8z};
  }
}

\newcommand\orcid[1]{\!%
  \href{https://orcid.org/#1}{%
    \mbox{%
      \scaleto{%
        \begin{tikzpicture}[yscale=-1,transform shape]
          \pic{orcidlogo};
        \end{tikzpicture}
      }{8pt}%
    }%
  }%
}
\begin{document}
\title{Quantum typicality approach to energy flow between two spin-chain domains
at different temperatures}

\author{Laurenz Beckemeyer~\orcid{0009-0004-2963-8689}}
\email{lbeckemeyer@uos.de}
\affiliation{University of Osnabr{\"u}ck, Department of Mathematics/Computer
Science/Physics, D-49076 Osnabr{\"u}ck, Germany}

\author{Markus Kraft~\orcid{0009-0008-4711-5549}}
\affiliation{University of Osnabr{\"u}ck, Department of Mathematics/Computer
Science/Physics, D-49076 Osnabr{\"u}ck, Germany}

\author{Mariel Kempa~\orcid{0009-0006-0862-4223}}
\affiliation{University of Osnabr{\"u}ck, Department of Mathematics/Computer
Science/Physics, D-49076 Osnabr{\"u}ck, Germany}

\author{Dirk Schuricht~\orcid{0000-0002-5255-780X}}
\affiliation{Institute for Theoretical Physics, Utrecht University, 3584CC
Utrecht, The Netherlands}

\author{Robin Steinigeweg~\orcid{0000-0003-0608-0884}}
\email{rsteinig@uos.de}
\affiliation{University of Osnabr{\"u}ck, Department of Mathematics/Computer
Science/Physics, D-49076 Osnabr{\"u}ck, Germany}

\date{\today}

%-------------------------------------------------------------------------------
% Abstract
%-------------------------------------------------------------------------------

\begin{abstract}
We discuss a quantum typicality approach to examine systems composed of two
subsystems at different temperatures. While dynamical quantum typicality is
usually used to simulate high-temperature dynamics, we also investigate
low-temperature dynamics using the method.
To test our method, we investigate the energy current between subsystems at
different temperatures in various paradigmatic spin-1/2 chains, specifically the
XX chain, the critical transverse-field Ising chain, and the XXZ chain. We
compare our numerics to existing analytical results and find a convincing
agreement for the energy current in the steady state for all considered models
and temperatures.
\end{abstract}

\maketitle

%-------------------------------------------------------------------------------
% Introduction
%-------------------------------------------------------------------------------
\section{Introduction}
Over the last decades, the broad field of quantum many-body systems out of
equilibrium has crystallized as one of the central topics of modern physics,
both experimentally and theoretically \cite{Bloch2008,
Polkovnikov2011, Eisert2015, Dalessio2016, Abanin2019}, ranging from fundamental
questions in statistical mechanics to applied questions in material science. The
understanding of nonequilibrium physics has developed remarkably over the last
years due to experimental advances \cite{Bloch2008}, new theoretical concepts
such as the typicality of pure states \cite{Alben1975,DeRaedt1989,
Jaklic1994,Hams2000,Gemmer2003,Iitaka2003,Gemmer2004, Goldstein2006,
Popescu2006, Reimann2007,Bartsch2009, Sugiura2012,Sugiura2013,Elsayed2013,
Steinigeweg2014,Monnai2014,Endo2018,Richter2019, Wietek2019, Rousochatzakis2019,
Heitmann2020,Schnack2020, Jin2021, Reimann2018, Mitric2025} and
eigenstate thermalization \cite{Deutsch1991, Srednicki1994, Rigol2008}, and the
development of sophisticated numerical techniques \cite{Jaklic2000,Long2003,
Schollwoeck2005,  Grossjohann2010,Schollwoeck2011,DeRaedt2017}.

One paradigmatic example of a nonequilibrium process is transport
\cite{Bertini2021}, a natural phenomenon in systems with one or more conserved
quantities. Additionally, transport appears in both open and closed systems. In
an open system scenario, transport processes are usually induced by coupling
the system  at its boundaries to external baths at different temperatures or
chemical potentials. Then, in the long-time limit, a nonequilibrium steady
state usually emerges with a characteristic density profile and a constant
current  \cite{Michel2003, Prosen2009, Znidaric2011,Xu2023}. Such a situation is
often modeled by a quantum master equation of Lindblad form \cite{Breuer2007},
which enables the application of sophisticated numerical techniques.

In a closed system scenario, a widely used approach for studying transport is
linear response theory, where the Kubo formula in terms of autocorrelation
functions plays a crucial role \cite{Kubo1991}. This theory predicts the
behavior of a system near  equilibrium, and it can be formulated either in
frequency and momentum space or in the space and time
domain. Another approach for studying transport in closed systems is a quantum
quench \cite{Calabrese2006, Mitra2018,Richter2018a}. Here, a
quantum system is
prepared in a particular initial state, typically
an eigenstate of some pre-quench Hamiltonian. Then, the system undergoes a
sudden change  and is described by a new Hamiltonian, and the initial
state evolves unitarily according to this post-quench Hamiltonian. Quantum
quenches can be studied in bipartite systems \cite{Bertini2016, Fischer2020},
where the total system is composed of two subsystems and initially
both subsystems are prepared in a state according to their respective
Hamiltonians. Then, the subsystems are brought into contact together, and the
total state evolves under the Hamiltonian of the total system.

Numerically, transport in open and closed systems has been
investigated using various methods, including exact and Lanczos
diagonalization \cite{Jaklic2000,Long2003, Karrasch2013}, simulations based
on matrix product states \cite{Vidal2004,Zwolak2004,Verstraete2008,
Prosen2009, Schollwoeck2011, Weimer2021,Paeckel2019}, and
Monte-Carlo techniques
\cite{Alvarez2002,Michel2008,Grossjohann2010,DeRaedt2017}, to name a few.
Additionally, the method of dynamical quantum typicality
\cite{Heitmann2020, Jin2021} is a multifaceted method for
investigating transport in
closed quantum systems. The basic idea of dynamical quantum typicality is that
one can imitate the expectation value of an entire ensemble by an expectation
value of a single pure state, drawn at random from a high-dimensional Hilbert
space. Dynamical quantum typicality has been used for
linear-response functions, mostly at high temperatures \cite{Bertini2021}. For
such temperatures,
the statistical error is smaller than at low temperatures. Still, dynamical
quantum typicality is a priori neither restricted
to linear-response functions nor high temperatures.

In this work, we demonstrate the usefulness of dynamical quantum typicality in
a specific physical situation, which differs from most previous applications in
two ways.
First, we apply it to energy flow in bipartite systems.  Second, we also explore
low temperatures. We do so for various spin-1/2 chains, which all feature
ballistic energy transport: the XX chain, the critical transverse-field Ising
chain, and the XXZ chain. We compare our numerical results to existing
analytical results from conformal field theory (CFT) \cite{Bernard2012,
Fischer2020} and generalized hydrodynamics (GHD) \cite{Bertini2016,
Castro-Alvaredo2016, Doyon2025}. We find convincing agreement  in all considered
subsystem configurations of our bipartite setup, as well as for all examined
temperatures. This agreement is not only found for the contact current but also
for all local currents and densities in the steady state.

Our paper is structured as follows: In Sec. \ref{sec:setup}, we define the
bipartite setup used throughout the paper and the energy current between the two
subsystems. After that, we introduce the different spin-$1/2$
chains in Sec. \ref{sec:models}, which we use for our subsystem configurations.
Next, in Sec. \ref{sec:methods}, we discuss our quantum typicality approach to
the energy current and briefly review the CFT results.
Thereafter, we present our numerical results and compare them to analytical
results in Sec.\ \ref{sec:results}. We conclude in Sec.\
\ref{sec:conclusion} and give directions for further research.
%-------------------------------------------------------------------------------
% Setup
%-------------------------------------------------------------------------------
\section{Setup} \label{sec:setup}

%-------------------------------------------------------------------------------
\begin{figure}[t]
\includegraphics[width=0.9\columnwidth]{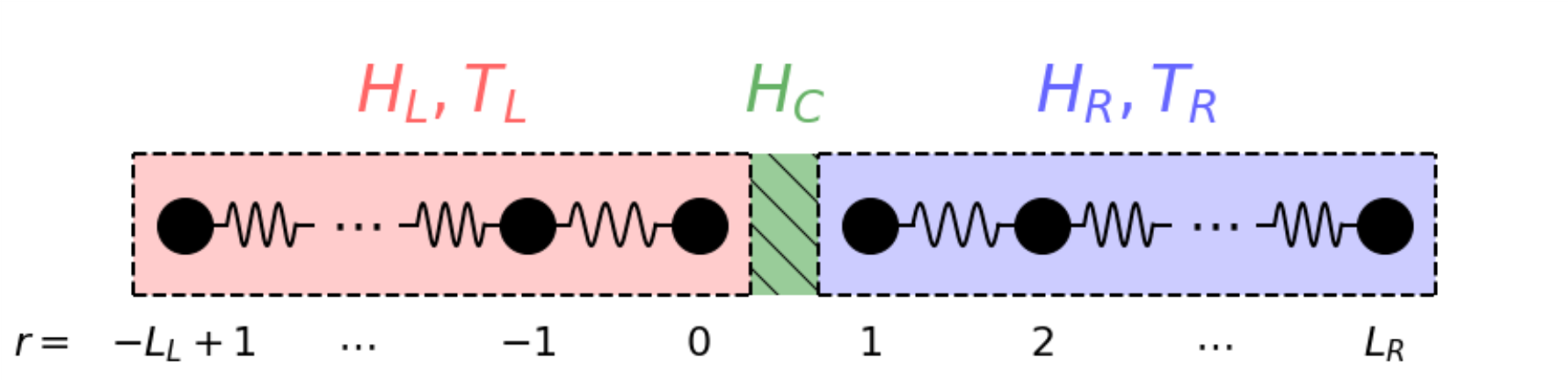}
\caption{Schematic representation of the setup. The total system
consists of two subsystems described by $H_{L}$ and $H_{R}$, respectively.
Initially, they are prepared at temperatures $T_L$ and $T_R$. The
interaction of both parts for times $tJ \geq 0$ is
described by the Hamiltonian $H_C$.}
\label{fig:sketch}
\end{figure}
%-------------------------------------------------------------------------------

Throughout our paper, we consider a system comprising two subsystems. Thus, the
full Hamiltonian is written as
\begin{eqnarray}
H = H_{L} + H_{R} + H_{C}\; ,
\label{eq:H_total_sys}
\end{eqnarray}
where $H_L$ and $H_{R}$ are the Hamiltonian of the left subsystem and right
subsystem, respectively. The term $H_{C}$ describes the interaction
between both subsystems. By $L_{L}$ and $L_{R}$ we denote the number of sites
of the two subsystems such that $L = L_{L} + L_{R}$ is the total number of
sites of the system. Here, we only focus on the equally sized case of $L_{L} =
L_{R} = L/2$. A schematic representation of our setup is shown in Fig.\
\ref{fig:sketch}.

In our paper, we are interested in the heat current from the left half into the
right one. This heat current is given by
\cite{Giazotto2006}
\begin{eqnarray}
Q_{L} = J_{0} - \mu \dot{N}_{L} \; ,
\end{eqnarray}
where $J_{0} = - \dot{E}_{L}$ denotes the energy current flowing out of the left
subsystem, $\dot{N}_{L}$ is the particle current, and $\mu$ is a chemical
potential. We focus on the case of $\mu = 0$. Thus, the heat current equals the
energy current, $Q_{L} = J_{0}$. By using the Heisenberg equation of motion
$(\hbar = 1)$,
we can express the energy current $J_{0}$ as
\begin{eqnarray}
J_{0} = -\dot{E}_{L} = - \frac{\text{d}}{\text{d}t}\langle H_{L} \rangle = i
\langle
[H_{L},H]\rangle \; ,
\label{eq:energy_current_J0}
\end{eqnarray}
where the brackets denote the expectation value with respect to some state.

Since we are interested in the energy current $J_{0}$, we
prepare both subsystems initially in a canonical ensemble with
different low temperatures $T_{L}$ and $T_{R}$ with respect to the Hamiltonian
$H_{L}$ and $H_{R}$. Thus, we have the canonical states
\begin{eqnarray}
\rho_{\beta_{k}}=\exp(-\beta_{k} H_{k})/Z,
\end{eqnarray}
where $k=L,R$ denotes the subsystem, $\beta_{k}=1/T_{k}$ the inverse temperature
and $Z_{k}=\text{tr}[\exp(-\beta_{k} H_{k})]$ the respective partition function.
Here, we have set $k_{B}=1$.

We investigate the energy current for various subsystem configurations,
including the coupling of (i)  two XX chains, (ii) two critical Ising chains,
(iii) one XX chain to a critical Ising chain, and (iv) two XXZ chains. We
introduce all models in the following.

%-------------------------------------------------------------------------------
\begin{figure}[t]
\includegraphics[width=0.9\columnwidth]{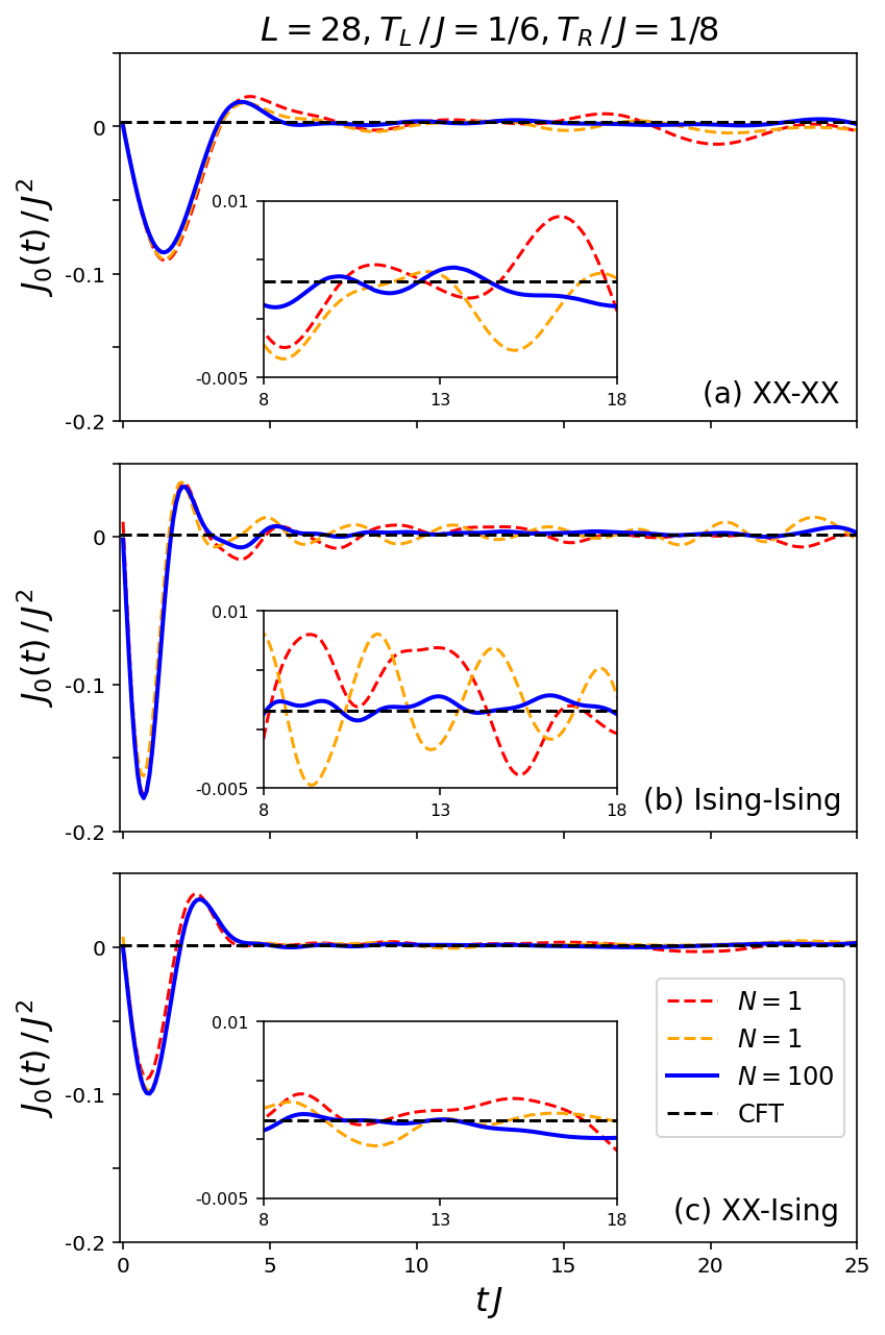}
\caption{Time evolution of the energy current for temperatures $T_{L}/J=1/6,
T_{R}/J=1/8$, total system size $L=28$ for the coupling of (a) two XX chains,
(b) two critical Ising chains and (c) one XX chain to a critical Ising chain.
Numerical data for two different pure random initial states ($N$=1) and for the
average over many initial states (N=100) are compared to the CFT value.
Note that larger fluctuations at longer times do not
contradict the bound in Eq.\ (\ref{eq:bound}), since this bound is not tight.}
\label{fig:time_evo}
\end{figure}
%-------------------------------------------------------------------------------

%-------------------------------------------------------------------------------
% Models
%-------------------------------------------------------------------------------
\section{Models} \label{sec:models}
In our paper, we examine three different examples of spin-1/2 models, which
correspond either to noninteracting systems or interacting integrable systems.
All of them feature ballistic transport of energy.  For simplicity, we focus on
the homogeneous case and comment on the subsystem configuration afterwards. All
models are one-dimensional lattice models of
the type
\begin{eqnarray}
H = \sum_{r=1}^{L-1} h_{r}\; ,
\end{eqnarray}
where $h_{r}$ are local Hamiltonians and $L$ denotes  again the number of
lattice
sites.

First, we introduce the spin-$1/2$ XXZ model. For this model, the local
Hamiltonians are given by \cite{Bertini2021}
\begin{eqnarray}
h_{\text{XXZ},r} =
 J\left(S_{r}^{x}S_{r+1}^{x}+S_{r}^{y}S_{r+1}^{y}+\Delta
S_{r}^{z}S_{r+1}^{z}\right) \; ,
\label{eq:H_XXZ}
\end{eqnarray}
where $S_{r}^{i}\ (i=x,y,z)$ are the spin-$1/2$ operators at site $r$, $J > 0$
is the antiferromagnetic exchange coupling constant, and $\Delta$ is the
anisotropy in $z$ direction.
Using the Jordan-Wigner transformation \cite{Jordan1928}, the spin-$1/2$ XXZ
chain can be mapped onto a chain of spinless fermions. After this
transformation, the local Hamiltonians in Eq.\ (\ref{eq:H_XXZ}) take on the form
\begin{eqnarray}
h_{\text{XXZ},r}  =   & &\frac{J}{2} \big(c_{r}^{\dagger}c_{r+1}
 +\text{h.c.} \big)  \\ \nonumber
 &+& J \Delta \left(n_{r} - \frac{1}{2} \right) \left(
n_{r} - \frac{1}{2} \right) \; ,
\label{eq:H_XXZ_trafo}
\end{eqnarray}
where $c_{r}^{\dagger}$ and $c_{r}$ are fermionic creation and annihilation
operators at site $r$, and $n_{r} =c_{r}^{\dagger}c_{r}$ denotes the
corresponding occupation number operator.

%-------------------------------------------------------------------------------
\begin{figure}[t]
\includegraphics[width=0.9\columnwidth]{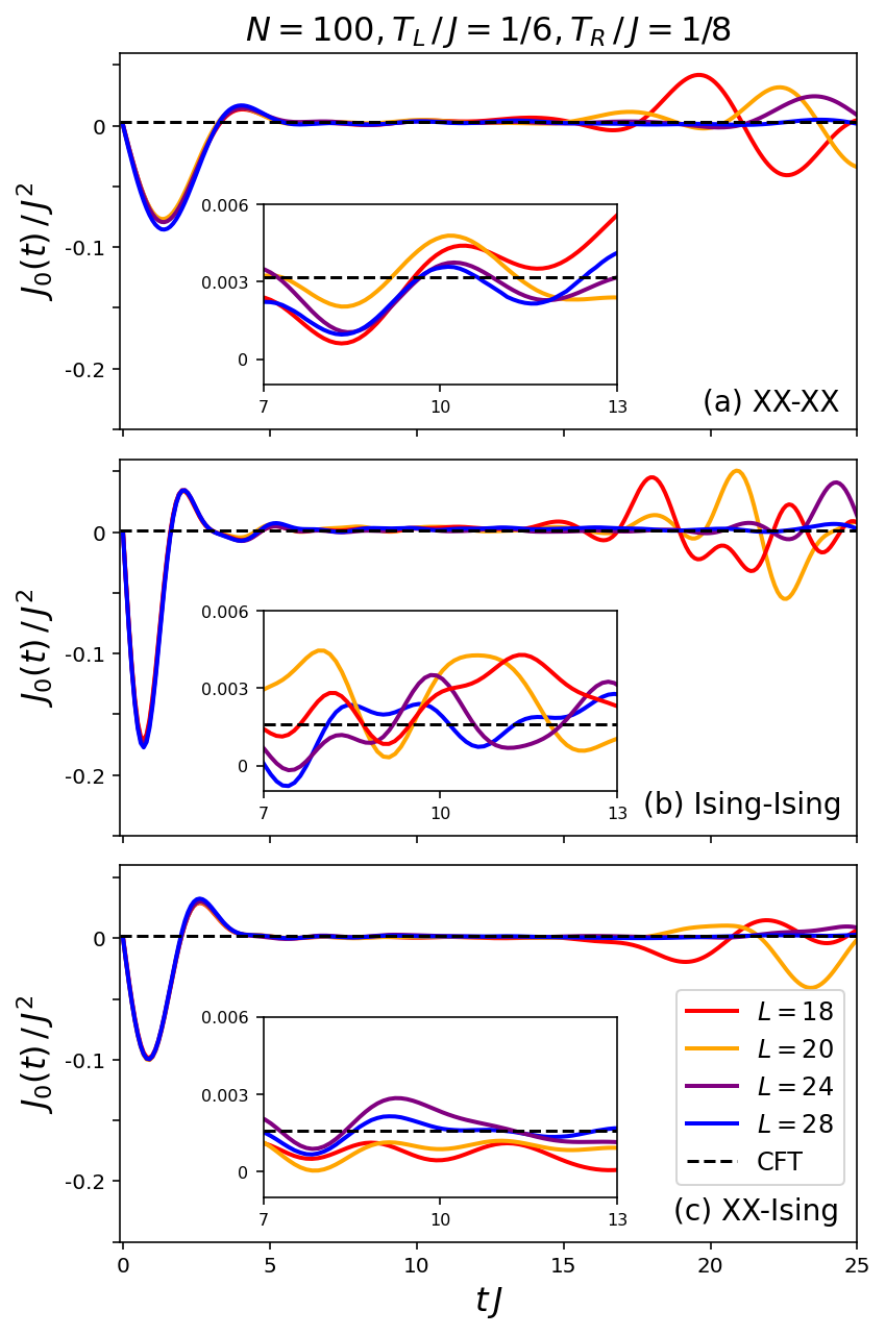}
\caption{Similar data as the one in Fig.\ \ref{fig:time_evo} but now for
different total system sizes $L$. Here, the numerical data are again averaged
over $N = 100$ pure random initial states. For larger system sizes $L$, finite
size effects are visible for later times $tJ$.}
\label{fig:system_size}
\end{figure}
%-------------------------------------------------------------------------------

The second model we consider is the special case of the spin-$1/2$ XXZ chain
with $\Delta = 0$, i.e., the XX chain. The local Hamiltonian reads
\begin{eqnarray}
h_{\text{XX},r} = J \left(S_{r}^{x}S_{r+1}^{x}+S_{r}^{y}S_{r+1}^{y} \right ) \;
.
\label{eq:H_XX}
\end{eqnarray}
This model is equivalent to a free-fermion chain, as can be seen by setting
$\Delta = 0$ in Eq.\ (\ref{eq:H_XXZ_trafo}) above.

The last  spin model we investigate in this paper is the
transverse-field Ising model. This model is described by the local
Hamiltonian
\begin{eqnarray}
h_{\text{Ising},r} = J  S_{r}^{x}S_{r+1}^{x} + \frac{h}{2}
(S_{r}^{z}+S_{r+1}^{z})  \; ,
\label{eq:H_Ising}
\end{eqnarray}
where the constant $h$ denotes the strength of the transverse magnetic field.
Due to the Jordan-Wigner transformation, the transverse-field Ising model is,
up to a constant, equivalent to the Kitaev chain
\cite{Lieb1961,Pfeuty1970,Kitaev2001,Suzuki2013},
\begin{eqnarray}
h_{\text{Kitaev},r} = \frac{J}{4} \Big(c_{r}^{\dagger} -
c_{r}\Big) \Big( c_{r+1}^{\dagger}+c_{r+1} \Big)
+ h n_{r} \; .
\end{eqnarray}
Throughout our paper, we focus on the critical transverse-field Ising
chain. Therefore, we set $h=1/2$.

As mentioned earlier, we investigate total systems with $L$ sites, composed of
two subsystems with $L_{L}=L_{R}=L/2$ sites each. As indicated in Fig.\
\ref{fig:sketch}, we slightly change notation and use for the left subsystem the
sites $r=- L_{L}+1,\dots, 0$ and for the right system the sites $r=1,\dots,
L_{R}$. The left subsystem interacts with the right subsystem at the sites $r=0$
(left half) and $r=1$ (right half). The interaction is described by the
Hamiltonian $H_{C}$, which changes for our different configurations.
We use  in the case of (i) the coupling of two XX chains
\begin{eqnarray}
H_{C} = J_{C} (S_{0}^{x}S_{1}^{x} + S_{0}^{y}S_{1}^{y}) \; ,
\end{eqnarray}
for the case of the coupling of (ii) two critical Ising chains and  (iii) one XX
chain to a critical Ising chain the term
\begin{eqnarray}
H_{C} = J_{C}S_{0}^{x}S_{1}^{x} \; ,
\end{eqnarray}
and for the coupling of (iv) two XXZ chains the term
\begin{eqnarray}
H_{C} = J_{C}\left(S_{0}^{x}S_{1}^{x}+S_{0}^{y}S_{1}^{y}+\Delta_{C}
S_{0}^{z}S_{1}^{z}\right)\; .
\end{eqnarray}
In our paper, we set $J_{C} = 1$ for all configurations. All chosen coupling
Hamiltonians ensure that energy can flow between both subsystems. Before we
proceed to our numerical results for the energy current in the various
subsystem configurations, we introduce our dynamical typicality approach. We
also briefly discuss the result obtained within CFT \cite{Bernard2012}.
%-------------------------------------------------------------------------------
% Methods
%-------------------------------------------------------------------------------
\section{Methods} \label{sec:methods}

%-------------------------------------------------------------------------------
% Dynamical typicality
%-------------------------------------------------------------------------------
\subsection{Dynamical typicality }

%-------------------------------------------------------------------------------
\begin{figure}[t]
\includegraphics[width=0.9\columnwidth]{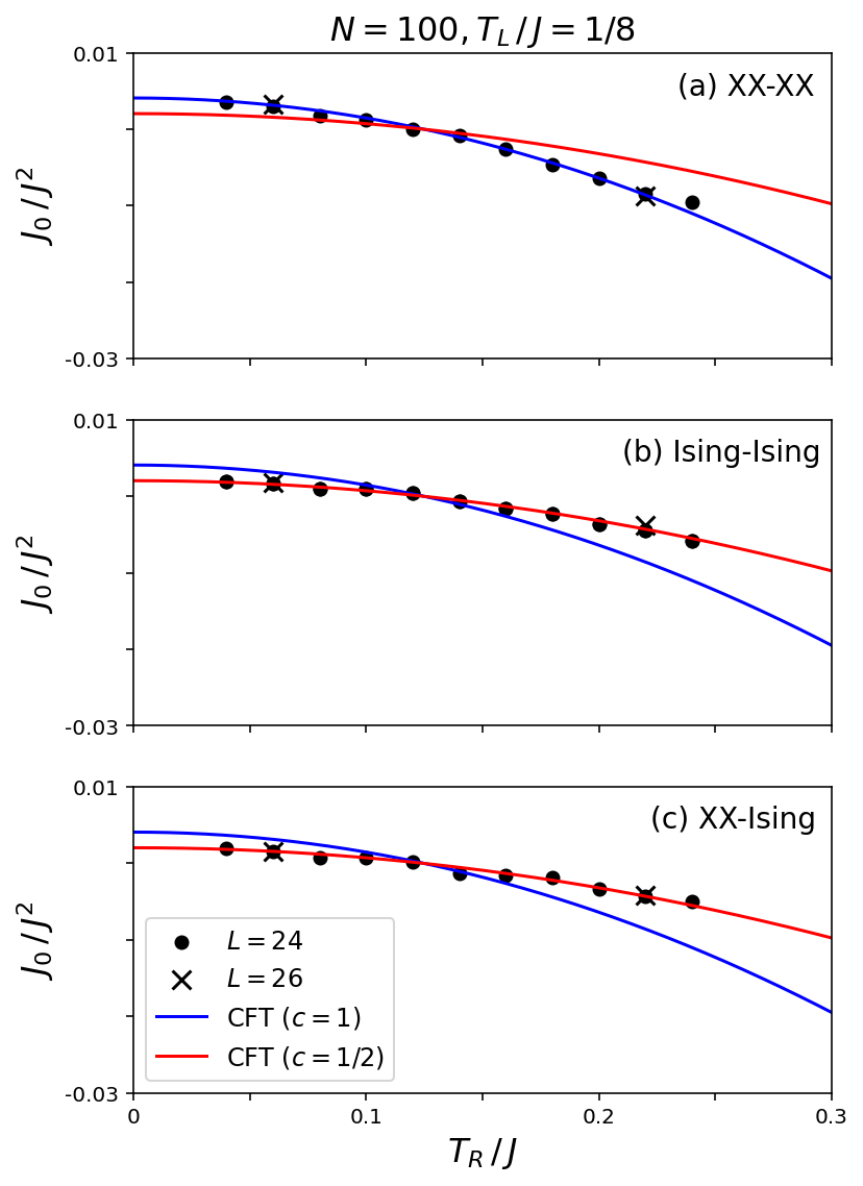}
\caption{Steady-state energy current $J_{0}$ versus temperature of
the right subsystem $T_{R}$  for the coupling of (a) two XX chains, (b)
two critical Ising chains, and (c) one XX chain to a critical Ising chain for
the total system size $L=24$ and $T_{L}/J = 1/8$. The numerical data are
compared to the CFT result (\ref{eq:energy_cft}).}
\label{fig:current}
\end{figure}
%-------------------------------------------------------------------------------

We now turn to a discussion of dynamical quantum typicality (DQT)
\cite{Jin2021,Heitmann2020}, which is the
key concept used in our numerical approach. We first focus on the homogeneous case
of a single Hamiltonian $H$ and a single temperature $T$, and afterwards come to
the bipartite case with two $H_L$, $H_R$ and two $T_L$, $T_R$.

The basic idea of DQT \cite{Gemmer2003,Goldstein2006,Popescu2006} is that one
can approximate the expectation value of
an ensemble density matrix,
\begin{equation}
\text{tr}[\rho(t) O] \, , \quad \rho(t) = e^{-iHt}\, \rho \, e^{iHt} \, ,
\end{equation}
by an expectation value of just one pure state,
\begin{equation}
\langle \psi(t) | O | \psi(t) \rangle \, , \quad | \psi(t) \rangle= e^{-i Ht}
\, | \psi
\rangle \, ,
\end{equation}
where $O$ is a local operator. For such a local operator,
the moments
\begin{equation}
M_n = \frac{\text{tr}[ O^n ]}{D}
\end{equation}
with the Hilbert-space dimension $D$, do not depend on system size $L$.
Moreover, we assume that these moments are finite.

If $\rho$ was the microcanonical ensemble and $|
\psi \rangle$ an energy eigenstate, this idea would be identical to the
well-known eigenstate thermalization hypothesis (ETH) \cite{Deutsch1991,
Srednicki1994, Rigol2008}. In contrast to ETH,
however, DQT is no assumption and relies on a particular construction of the
pure state $| \psi \rangle$. Specifically, for a given ensemble density
matrix $\rho$, this construction reads
\begin{equation} \label{eq:construction}
| \psi \rangle  = | \psi (0) \rangle \propto \sqrt{\rho} \, | \Phi \rangle
\, ,
\end{equation}
which is the action of the square-root of $\rho$ on a so-called Haar-random
pure state $ | \Phi \rangle$ \cite{Bartsch2009}. As
discussed below, this construction particularly enables the reduction of memory
in numerical simulations, since pure states (of dimension $D$) instead of
density matrices (of dimension $D^2$) can be used.
\\
The particular pure state $| \Phi \rangle$ does neither depend on $\rho$ nor
$H$. In an arbitrarily chosen orthogonal basis, it is given by
\begin{equation}
| \Phi \rangle \propto \sum_{m=1}^{D} c_{m} | m \rangle \, ,
\end{equation}
where $D$ is again the Hilbert-space dimension and $c_m$
are complex
coefficients. Both, the real and imaginary part of these coefficients
are drawn at random according to a Gaussian probability distribution with zero
mean. For such a $| \Phi \rangle$ and the resulting $| \psi \rangle$, DQT then
leads to the approximation \cite{Reimann2007, Bartsch2009}
\begin{equation} \label{eq:approximation}
\text{tr}[\rho(t) O] = \frac{\langle \psi(t) | O | \psi(t) \rangle}{\langle
\psi | \psi \rangle} + \varepsilon \, ,
\end{equation}
where the error $\varepsilon$ depends on the particular
realization of the random state $| \Phi \rangle$. However, it possible to make a
statistical statement and $\varepsilon$ vanishes on average. More
importantly, however, the standard deviation is bounded from above,
\begin{equation} \label{eq:bound}
\sigma(\varepsilon) \leq b = {\cal O}\left(\frac{1}{\sqrt{D_\text{eff}}}\right)
\, .
\end{equation}
Here, $D_\text{eff}$ is the so-called effective dimension. In the case of the
canonical ensemble $\rho \propto \exp (-\beta H)$, which is most relevant in the
context of our work and explained in more detail in Appendix
\ref{sec:details}, the effective dimension is given by the partition function
\begin{equation}
D_\text{eff} = \text{tr}[\exp (-\beta (H - E_0))]
\end{equation}
with the ground-state energy $E_0$. Thus, $D_\text{eff}$ counts the number of
thermally occupied energy eigenstates. In the high-temperature limit $T \to
\infty$, $D_\text{eff} = D = 2^L$ and hence the statistical error
$\varepsilon$ decreases exponentially fast with system size $L$. In the
zero-temperature limit $T\to 0$, however, $D_\text{eff} = 1$ and thus the
statistical error $\varepsilon$ can be in principle large.
Still, one has to
take into account that the bound in Eq.\ (\ref{eq:bound}) is not tight. In
fact, at $T = 0$, it turns out that the statistical error becomes $\varepsilon
= 0$. This fact is a kind of trivial consequence for a Haar-random pure state
$| \Phi \rangle$: It is also random in the energy eigenbasis and the
coefficient $c_0$ for the ground state $| 0 \rangle$ might be small but
likely nonzero. At $T = 0$, only this coefficient survives and one has $| \psi
\rangle = | 0 \rangle$ \cite{Balz2018}. Note that this fact
is important, since we have $D_\text{eff}$ close to 1 for the low temperatures
considered later, in contrast to the large $D_\text{eff}$ in usual
high-temperature consideratios, e.g., $D_\text{eff} = D = 2^{30}$ for
$L=30$.

Let us next turn to the specific setup in Fig.\ \ref{fig:sketch}, which is
studied in our work. For this setup, we could proceed completely analogous and
choose for the construction of the pure state $| \psi \rangle$ in Eq.\
(\ref{eq:construction}) an ensemble density matrix of the form
\begin{equation}
\rho \propto \exp (-\beta_L H_L) \otimes \exp (-\beta_R H_R) \, .
\end{equation}
While this choice is valid, the resulting $| \psi \rangle$ would be no product
state of the two halves of the system, because $| \Phi \rangle$ is drawn at
random from the full Hilbert space and thus highly entangled. Hence, to
implement the product structure, we proceed slightly different and choose the
pure state as
\begin{equation}
| \psi \rangle \propto \exp \left(-\frac{\beta_L H_L}{2}\right) | \Phi_L
\rangle \otimes
\exp \left(-\frac{\beta_R H_R}{2}\right) | \Phi_R \rangle \, ,
\end{equation}
where $| \Phi_L \rangle$ and $| \Phi_R \rangle$ are now drawn at random from the
Hilbert space of the left and right half. While this choice
does not change the typicality relation in Eq.\ (\ref{eq:approximation}),
it
goes along with a larger statistical error $\varepsilon$, which is given by
the partition functions of the two halves, rather than the partition function of
the entire system. To further reduce the statistical error, we average over $N =
100$ different realizations of $| \psi \rangle$,
\begin{equation} \label{eq:approximation}
\text{tr}[\rho(t) O] = \frac{1}{N} \sum_{i=1}^N \frac{\langle \psi_i(t) | O |
\psi_i(t) \rangle}{\langle \psi_i | \psi_i \rangle} \, ,
\end{equation}
if not stated otherwise. In this way, the statistical
error in Eq.\ (\ref{eq:bound}) is multiplied by the factor $1/\sqrt{L}$
\cite{Jaklic2000}. In our case, the local observable $O$ is mostly the
contact current $J_0$, but we also consider currents $J_r$ at other sites $r$ as
well as the local energy densities $h_r$.

In our numerical simulations, we calculate the action of the
imaginary-time  operators $e^{-\beta_L H_L/2}$ and $e^{-\beta_R H_R/2}$
on the Haar-random pure states $| \Phi_L \rangle$ and $| \Phi_R \rangle$ with a
fourth-order Runge-Kutta scheme, where we choose a very small step
$\delta \beta J = 0.0001$. This scheme is also used for the action of the
real-time  operator $e^{-i H t}$ on the pure state $| \psi \rangle$, where we
choose a larger but still small step $\delta t J = 0.01$.
By doing so and also using sparse-matrix techniques, we can
treat system sizes outside the range of standard exact diagonalization.
Alternatively, other propagation methods, such as Trotter decompositions
\cite{Suzuki1985, DeRaedt1987, Berry2006} or Chebyshev polynomial expansions
\cite{Tal-Ezer1984,Weise2006}, can be used as well.

%-------------------------------------------------------------------------------
% conformal field theory
%-------------------------------------------------------------------------------
\subsection{Conformal field theory}

%-------------------------------------------------------------------------------
\begin{figure}[t]
\includegraphics[width=0.9\columnwidth]{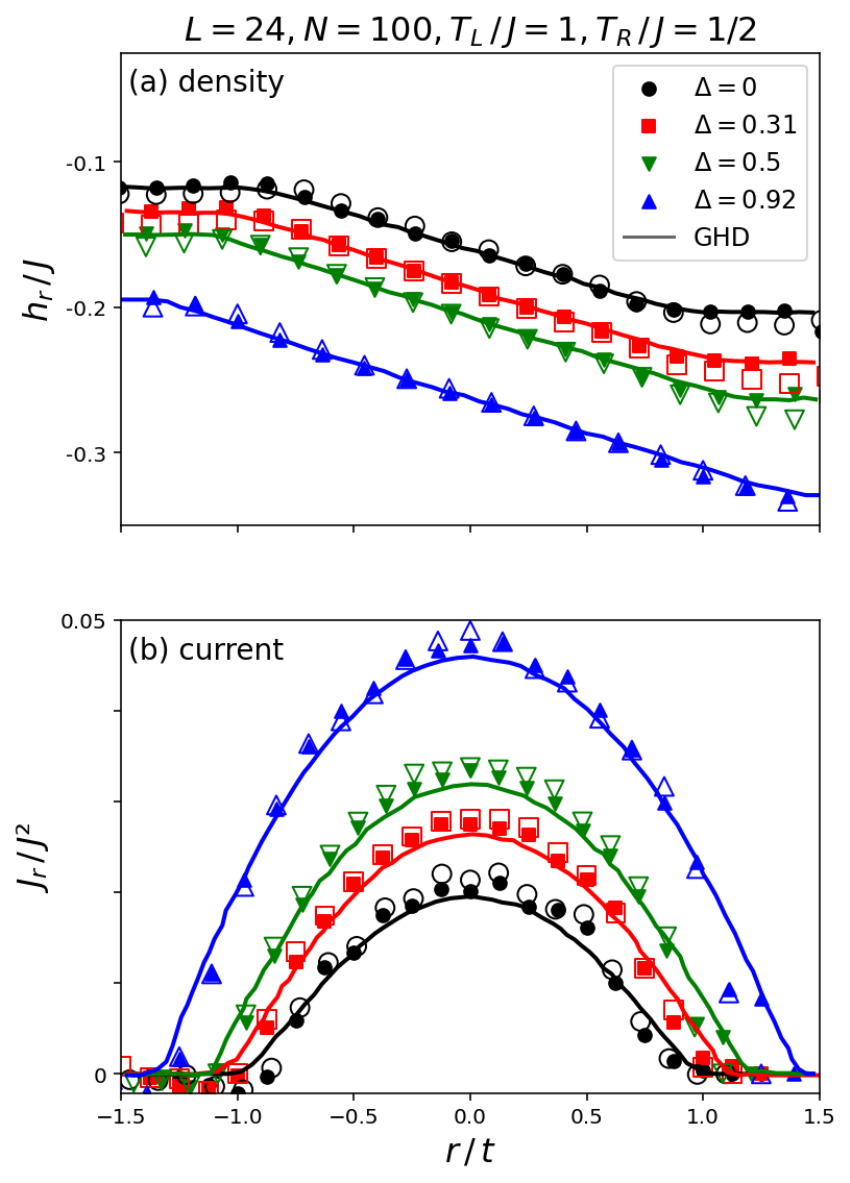}
\caption{Steady-state profile for (a) the local density $\langle h_{r}\rangle$
and (b) the local current $\langle J_{r}\rangle$ for the coupling of two XXZ
chains with the same anisotropies $\Delta$, $T_{L}/J=1$,$T_{R}/J=1/2$, and
total system size $L=24$. (Note that in the $x$-axis the
position $r$ is divided by the time $t$.) The numerical data are compared to
the results from GHD in Ref.\ \cite{Bertini2016}. Here, the numerical data are
shown for both, a random product state (open symbols) and a random state in the
entire Hilbert space (closed symbols).}
\label{fig:GHD_XXZ}
\end{figure}

%-------------------------------------------------------------------------------
The energy transport in a bipartite CFT setup has been studied by Bernard and
Doyon \cite{Bernard2012}. Specifically they considered a CFT separated into two
halves prepared at different temperatures. These halves were then glued together
and the non-equilibrium steady state emerging at late times was studied. In
particular, for the energy current they obtained the universal formula
\begin{eqnarray}
J_E=\frac{\pi c}{12}(T_L^2-T_R^2) \; ,
\label{eq:energy_cft}
\end{eqnarray}
where $c$ denotes the central charge of the CFT. Given that CFTs generically
describe the universal low-energy behavior of certain lattice models, the
result (\ref{eq:energy_cft}) is expected to hold only in the low-temperature
regime. The central charge $c$ is a specific characteristic of the
CFT \cite{DiFrancesco1997}, physically it can be interpreted as the number of
degrees of freedom, which here contribute to the energy transport through the
system.

The setup leading to the universal CFT result (\ref{eq:energy_cft}) possesses
translation invariance after the gluing quench. As an extension,
Ref. \cite{Fischer2020} studied the coupling of an XX chain to a critical Ising
chain, i.e., our setup (iii) above. Using both the non-equilibrium Green
function formalism and the density matrix renormalization group method, the
energy current was found to behave as
\begin{eqnarray}
J_E=\frac{\pi}{12}\text{min}(c_L,c_R)\,(T_L^2-T_R^2) \; .
\label{eq:energy_cft_min}
\end{eqnarray}
This result can be viewed as a bottleneck effect, where the subsystem with the
lower number of degrees of freedom as reflected by the lower central charge
limits the energy transport. The result (\ref{eq:energy_cft_min}) is also
consistent with a purely field theoretic analysis by Bernard et
al. \cite{Bernard2015}, where, however, the precise link to a microscopic
setup remains elusive.

%-------------------------------------------------------------------------------
% Results
%-------------------------------------------------------------------------------
\section{Results} \label{sec:results}

%-------------------------------------------------------------------------------
\begin{figure}[t]
\includegraphics[width=0.9\columnwidth]{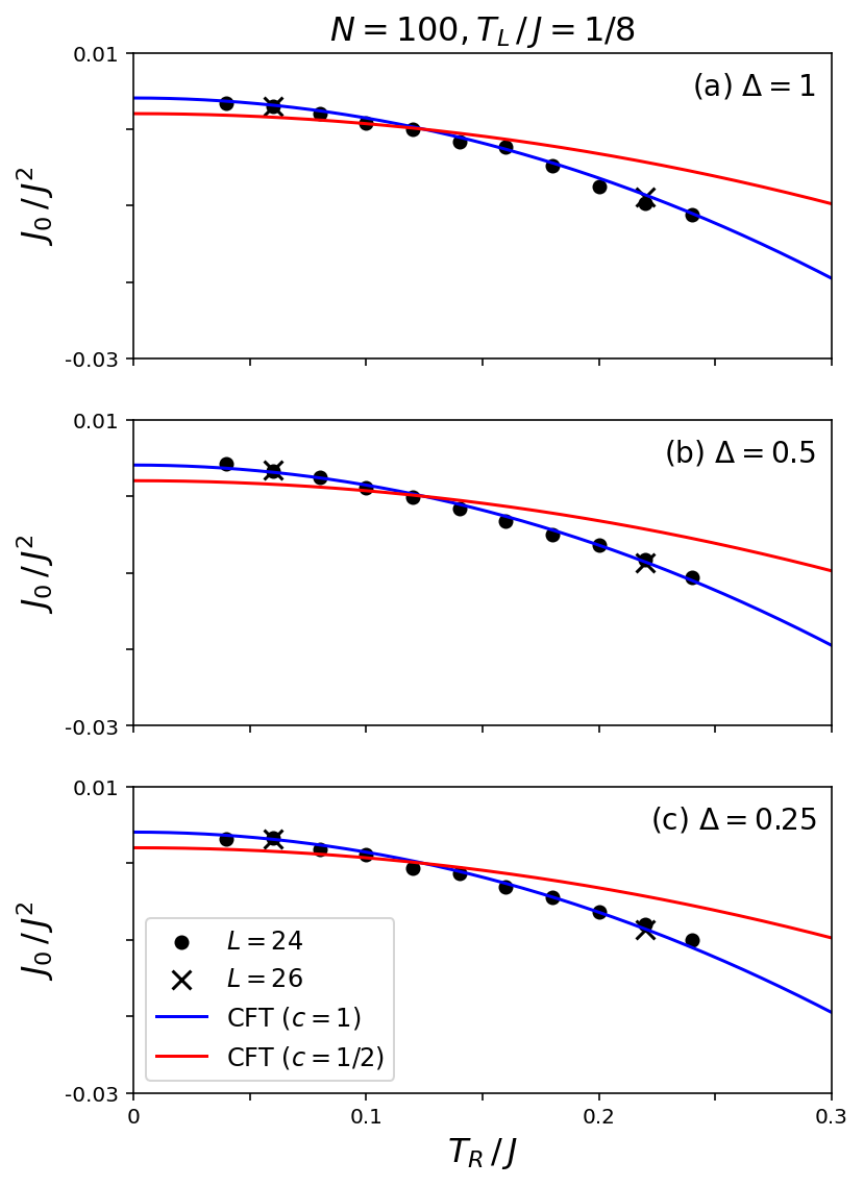}
\caption{Similar data as the one in Fig.\ \ref{fig:current} but now for the
coupling of two XXZ chains with anisotropies (a) $\Delta = 1$, (b) $\Delta=
0.5$, and (c) $\Delta = 0.25$.}
\label{fig:current_XXZ}
\end{figure}
%-------------------------------------------------------------------------------

Now, we turn to our numerical results, where we compare the energy current for
various subsystem configurations to the value from CFT in Eq.\
(\ref{eq:energy_cft}).
We first focus on the coupling of (i) two XX chains, (ii) two critical Ising
chains, and (iii) one XX chain to a critical Ising chain. In CFT, the XX chain
has the central charge $c=1$ while the transverse-field Ising chain at the
critical point has the central charge $c=1/2$ \cite{DiFrancesco1997}.

To verify our numerical method initially, we compute the time evolution of the
energy current $J_{0}$ for our considered subsystem configurations for systems
with $L=28$ sites and a fixed temperature of the left ($T_{L}/J=1/6$)
and right ($T_{R}/J = 1/8$) subsystem in Fig.\ \ref{fig:time_evo}.  All
subsystem configurations exhibit a transition regime characterized by
oscillations and a buildup of correlations across the interface. After a
specific timescale, the current reaches an approximately constant plateau,
indicating the establishment of a non-equilibrium steady state. Also in Fig.\
\ref{fig:time_evo}, we use two different realizations of a single random pure
state and compare them with the average over $N=100$  random pure
states. With the averaging, the statistical error reduces remarkably,
especially for the long-term behavior of the energy current $J_{0}$.
Furthermore, the steady state of the energy current agrees nicely with the
value from CFT for all subsystem configurations. In the following, we average
all our numerical results over $N=100$ random pure  initial states, if not
stated differently.

Next, we investigate the dependence of  the energy current on the total system
size. We thus depict the time evolution of the energy current $J_{0}$ for
varying total system sizes $L$ and for the fixed initial temperatures of
$T_{L}/J=1/6$ and $T_{R}/J = 1/8$ in Fig.\ \ref{fig:system_size}. For longer
times, finite-size effects are visible for smaller $L$. Nevertheless, below the
time where finite sizes effects are relevant, the data are close to the
steady-state value according to CFT, especially for the coupling of an XX chain
to the critical Ising chain, cf. Fig.\ \ref{fig:system_size}\ (c). In the
following, we focus on total system sizes of $L=24$, where finite-size effects
and deviations from the CFT value are sufficiently small.

As one of our central results, we now calculate the steady-state energy current
$J_{0}$ for a fixed initial temperature of $T_{L}/J = 1/8$ of the left
subsystem and different initial temperatures of the right subsystem, see  Fig.\
\ref{fig:current}. We compare our numerics to the CFT values in Eq.\
(\ref{eq:energy_cft}) for $c=1/2$ and $c=1$. For all considered subsystem
configurations (i) -- (iii), we have a convincing agreement between our
numerical results and  CFT.
Also our data do not significantly depend on system size, as we show in Fig.\
\ref{fig:current} by also calculating the energy current for exemplary
temperatures $T_{R}/J$ of the right subsystem for a total system size of
$L=26$.
Furthermore,  for the coupling of an XX chain to a critical Ising chain, the
energy current is determined by the system with the lower central charge, which
is, in this scenario, the critical Ising chain, see Fig.\ \ref{fig:current}\
(c). Consequently, we observe a bottleneck effect \cite{Fischer2020}, where the
energy transport is limited by the subsystem with the fewer degrees of freedom.

Let us also give a technical remark here. It would be
certainly interesting to explore even lower temperatures, e.g., to detect a
region, where eventually DQT and CFT differ significantly from each other.
However, in Fig.\ \ref{fig:current} we already consider a temperature $T_R/J$ =
0.05 or inverse temperature $\beta_R J= 20$. To do so, we have to accurately
approximate a small number of order $e^{-20}$ by an iterative procedure in
imaginary time. An accurate approximation for even lower temperatures is
quickly beyond the scope of such an interative procedure.

As the final subsystem configuration, we proceed to the (iv) coupling of two XXZ
chains. We first focus on the case where all anisotropies are equal, i.e.,
$\Delta = \Delta_{L} = \Delta_{R} = \Delta_{C}$. We again benchmark our
numerical calculation against existing results in the literature. In
Fig.\ \ref{fig:GHD_XXZ}, we compare our numerics for a total system with $L=24$
sites, fixed temperatures of the left and right subsystem, and different
anisotropies $\Delta$ to results from GHD. The numerical
data for the site dependence of the local energy density (Fig.\
\ref{fig:GHD_XXZ}\ (a)) as well as the data for the
site dependence of the local energy current (Fig.\ \ref{fig:GHD_XXZ}\ (b))
matches the results of Ref.\ \cite{Bertini2016} convincingly. Here, the
numerical data are shown for both, a random product state (open symbols) and a
random state in the entire Hilbert space (closed symbols). As shown in Fig.\
\ref{fig:GHD_XXZ}, the statistical error for a random state in the entire
Hilbert space is slightly smaller than the statistical error for a random
product state. It would be also interesting to analyze
states, which interpolate between these two states of no and maximum bipartite
entanglement. But we leave this analysis for future research.

Focusing again on the contact site, we calculate the energy current
$J_{0}$ for the coupling of two XXZ chains with total system size $L=24$, fixed
temperature $T_{L}/J=1/8$, and the anisotropies $\Delta=0.25,0.5$, and $1.0$,
see Fig.\ \ref{fig:current_XXZ}.  For $-1 \leq \Delta < 1$, the ground state of
the XXZ chain is gapless. The low-energy behavior of the XXZ chain in this
regime is described by a CFT with  central charge of $c=1$
\cite{Giamarchi2003}. Again, we recover the temperature
dependence in Eq.\ (\ref{eq:energy_cft}) for all considered anisotropies, and
our data agrees with the CFT result convincingly and also do not
significantly depend on system size.

\begin{figure}[t]
\includegraphics[width=0.9\columnwidth]{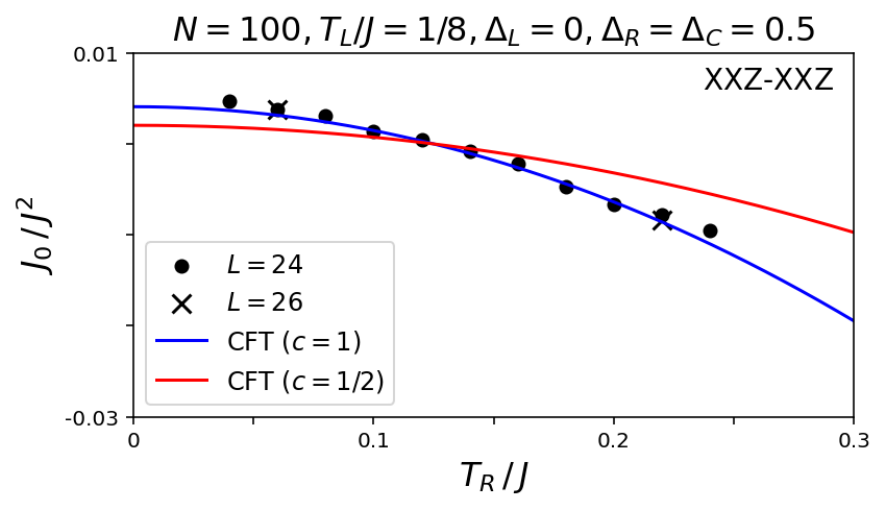}
\caption{Similar data as the one in Fig.\ \ref{fig:current_XXZ} but now for the
coupling of two XXZ chains with anisotropies $\Delta_{L} = 0$ and
$\Delta_{R}=\Delta_{C} = 0.5$.}
\label{fig:current_XXZ_different_D}
\end{figure}

At last, we also calculate the energy current $J_{0}$ for the coupling of two
XXZ chains, but now with the anisotropies $\Delta_{L} = 0$ and
$\Delta_{R}=\Delta_{C} = 0.5$. Additionally, in this case, we find convincing
agreement with the CFT value, as shown in Fig.\
\ref{fig:current_XXZ_different_D}. Overall, this and the previous results
demonstrate that our typicality approach is suitable to capture the energy
current down to low temperatures.

%-------------------------------------------------------------------------------
%-------------------------------------------------------------------------------

%-------------------------------------------------------------------------------
% Conclusion
%-------------------------------------------------------------------------------
\section{Conclusion} \label{sec:conclusion}

To summarize, we discussed the method of dynamical quantum typicality, mostly
used in linear response theory and at high temperatures, in less explored
contexts. First, we applied it to energy flow between systems
composed of two spin-chain domains at different  temperatures and,
second, to low temperatures. To test our method, we investigated the energy
current in different spin-$1/2$ subsystem configurations, namely the coupling
of  (i) two XX chains, (ii) two critical Ising chains, (iii) one XX chain to a
critical Ising chain, and (iv) two XXZ chains. We compared our numerical
results to results from CFT and GHD.
For all considered subsystem configurations and temperatures, we found a
convincing agreement between our quantum typicality approach and
these established methods. Therefore, we demonstrated the suitability of our
approach for the study of bipartite systems at low temperatures.

Interesting directions of future investigations include additional models,
e.g., the three-state
quantum Potts chain, which in CFT has a central charge of
$c=4/5$. In addition, it is desirable to extend the
investigation to gapped systems. Our expectation is that these systems are
harder to treat by means of DQT, compared to the critial ones studied here.
\\
%-------------------------------------------------------------------------------
% Acknowledgments
%-------------------------------------------------------------------------------
\section*{Acknowledgments}
This work was funded by the Deutsche Forschungsgemeinschaft (DFG)
under Grant No. 397067869 within the DFG Research Unit FOR 2692 under Grant No.
355031190.

%-------------------------------------------------------------------------------
% DATA AVAILABILITY
%-------------------------------------------------------------------------------

\section*{DATA AVAILABILITY}
The data that support the findings of this article are openly
available \cite{data}.

\appendix

\section{Details on the error bound}
\label{sec:details}

In the following, we provide details on the error bound for the
variance, which can be found in, e.g., \cite{Knipschild2021} in similar form.

Let us consider the variance
\begin{equation}
\sigma^2 = \overline{\langle \psi|O|\psi\rangle^{2}} -
(\overline{\langle \psi|O|\psi\rangle})^{2} \; ,
\end{equation}
where the overbar denotes the ensemble average over pure states $|\psi\rangle$.
For a pure state $|\psi\rangle$ according to Eq.\ (\ref{eq:construction}) one obtaines
the variance
\begin{eqnarray}\label{eq:sigma_long}
\sigma^2 &= & \sum_{i,j,k,l}\overline{c_{i}^{\ast}c_{j}c_{k}^{\ast}c_{l}}
\langle m_{i}|\sqrt{\rho} O\sqrt{\rho}|m_{j}\rangle  \langle m_{k}|\sqrt{\rho}
O\sqrt{\rho}|m_{l}\rangle  \nonumber \\
&& - \sum_{i,k} \langle m_{i}|\sqrt{\rho} O\sqrt{\rho}|m_{i}\rangle  \langle
m_{k}|\sqrt{\rho} O\sqrt{\rho}|m_{k}\rangle \; .
\end{eqnarray}
We now use the identity $\overline{c_{i}^{\ast}c_{j}c_{k}^{\ast}c_{l}} \approx
\delta_{ij}\delta_{kl}+\delta_{il}\delta_{jk}$. Here, we can neglect the case
$i=j=k=l$ for large dimensions $D \gg 1$ because this single case has only a
contribution of order $\mathcal{O}(1/D)$. With this identity, we can simplify
Eq. (\ref{eq:sigma_long}) to
\begin{eqnarray}
\sigma^{2} &\approx&  \sum_{i,j} \langle
m_{i}|\sqrt{\rho} O\sqrt{\rho}|m_{j}\rangle  \langle m_{j}|\sqrt{\rho}
O\sqrt{\rho}|m_{i}\rangle \;  \nonumber \\
&=& \text{tr}[\rho O \rho O] \, .
\end{eqnarray}
Next, we want to bound $\text{tr}[\rho O \rho O]$ from above.
To this end, we use the spectral decomposition $p =
\sum_{i} p_{i} |i \rangle \langle i |$ and get
\begin{eqnarray}
\text{tr}[\rho O \rho O] = \sum_{i,j}p_{i}p_{j} |\langle i | O|j\rangle|^{2}
\; .
\end{eqnarray}
Using
\begin{eqnarray}
(p_{i}-p_{j})^{2} = p_{i}^{2} - 2p_{i}p_{j} + p_{j}^{2} \geq 0,
\end{eqnarray}
we then obtain the inequality
\begin{eqnarray}
\text{tr}[\rho O \rho O] &\leq& \sum_{i,j} \frac{p_{i}^{2}+p_{j}^{2}}{2}
|\langle i | O|j\rangle|^{2} \nonumber \\
&=& \sum_{i} p_{i}^{2} \langle i|O^{2}|i\rangle \nonumber  \\ &=&
\text{tr}[\rho^{2}O^{2}] \; .
\end{eqnarray}
In the eigenbasis $|O_{j}\rangle$ of the operator $O$ we obtain for the trace
on the r.h.s the estimation
\begin{eqnarray}
\text{tr}[\rho^{2}O^{2}] &=& \sum_{j} \langle O_{j}|\rho^{2}|O_{j}\rangle\langle
O_{j}|O^{2}|O_{j}\rangle \nonumber \\
&\leq& O_\text{max}^2 \, \text{tr}[\rho^{2}] \; ,
\end{eqnarray}
where $O_\text{max} = \text{max}[O_j]$ is the maximal eigenvalue. Thus, in summary, the
variance can be bounded from above as
\begin{equation}
\sigma^2 \leq O_\text{max}^2 \, \text{tr}[\rho^2] \, ,
\label{eq:bound_sigma_2}
\end{equation}
where $O_\text{max}$ of the local operator $O$
does not depend on system size $L$. Inserting for the density matrix $\rho$ the
canonical ensemble leads to
\begin{equation}
\frac{\sigma^2}{O_\text{max}^2} \leq \frac{\text{tr}[(e^{-\beta
H})^2]}{(\text{tr}[e^{-\beta H}])^2}
\end{equation}
or, equivalently,
\begin{equation}
\frac{\sigma^2}{O_\text{max}^2} \leq \frac{\text{tr}[(e^{-\beta
(H - E_0)})^2]}{(\text{tr}[e^{-\beta (H - E_0)}])^2} =
\frac{\text{tr}[e^{-2 \beta
(H - E_0)}]}{(\text{tr}[e^{-\beta (H - E_0)}])^2}
\end{equation}
with the ground-state energy $E_0$.
Because
\begin{equation}
\text{tr}[e^{-2 \beta (H - E_0)}] \leq \text{tr}[e^{-\beta (H - E_0)}] \, ,
\end{equation}
one gets the expression
\begin{equation}
\frac{\sigma^2}{O_\text{max}^2} \leq \frac{\text{tr}[e^{-\beta
(H - E_0)}]}{(\text{tr}[e^{-\beta (H - E_0)}])^2} =
\frac{1}{\text{tr}[e^{-\beta (H - E_0)}]} \, .
\end{equation}
By introducing the effective dimension
\begin{equation}
D_\text{eff} = \text{tr}[e^{-\beta (H - E_0)}] \, ,
\end{equation}
this expression can be written as
\begin{equation}
\sigma^2 \leq O_\text{max}^2 \, \frac{1}{D_\text{eff}} \, .
\end{equation}
Omitting the $L$-independent $O_\text{max}$ finally yields
\begin{equation}
\sigma \leq b = {\cal O} \left( \frac{1}{\sqrt{D_\text{eff}}} \right ) \, ,
\end{equation}
cf., e.g., \cite{Jaklic2000}.
Note that this bound can be derived analogously for a time-dependent
operator $O(t)$ in the Heisenberg picture, because (i) $O(t)$ remains Hermitian under time evolution and (ii) the
spectrum and thus $O_{\text{max}}$ is invariant under time evolution.
Therefore, the bound in inquality (\ref{eq:bound_sigma_2}) holds for $O(t)$
with the same constant $O_{\text{max}}$.

%\newpage
%-------------------------------------------------------------------------------
% References
%-------------------------------------------------------------------------------
%
%\nocite{*}
%\bibliographystyle{apsrev4-1_titles}
%\bibliographystyle{apsrev4-2}
%\bibliography{Literature.bib}
%merlin.mbs apsrev4-1.bst 2010-07-25 4.21a (PWD, AO, DPC) hacked
%Control: key (0)
%Control: author (72) initials jnrlst
%Control: editor formatted (1) identically to author
%Control: production of article title (-1) disabled
%Control: page (0) single
%Control: year (1) truncated
%Control: production of eprint (0) enabled
%

\clearpage
\newpage
\end{document}